\documentstyle[aps,prl,epsf,floats,twocolumn]{revtex}

\def\JETPL#1#2#3{JETP Lett.\ {\bf {#1}}, {#2} (19{#3})}

\def\PRB#1#2#3{Phys.\ Rev.\ B {\bf {#1}}, {#2} (19{#3})}
\def\PRL#1#2#3{Phys.\ Rev.\ Lett.\ {\bf {#1}}, {#2} (19{#3})}

\def\SS#1#2#3{Surf.\ Sci.\ {\bf {#1}}, {#2} (19{#3})}

\def\ZhETF#1#2#3#4#5{Zh.E.T.F.\ {\bf {#1}}, {#2} (19{#3}) [JETP {\bf {#4}}, 
{#5} (19{#3})]}

\def\Nature#1#2#3{Nature {\bf {#1}}, {#2} (19{#3})}

\def\MicroE_Eng#1#2#3{Microelect.\ Eng.\ {\bf {#1}}, {#2} (19{#3})}

\def\SM#1#2#3{Superlatt.\ Microstr.\ {\bf {#1}}, {#2} (19{#3})}

\def\etal{{\it et al.}}
\def\Lphi{L_{\varphi}}
\def\dn{\delta_{n}}
\def\dnb{\delta_{n}^{*}}

\def\gap{\scriptsize${\stackrel{\textstyle _>}{_\sim}} $\normalsize \  }
\def\lap{\scriptsize${\stackrel{\textstyle _<}{_\sim}} $\normalsize \  }

\begin{document}
\draft
\twocolumn[\hsize\textwidth\columnwidth\hsize\csname @twocolumnfalse\endcsname
\title{Phase diagram and validity of one-parameter scaling 
near the two-dimensional metal-insulator transition}
\author{Nam-Jung Kim$^1$, Dragana Popovi\'{c}$^{2}$, and S.\ Washburn$^{1}$}
\address{$^{1}$Dept.\ of Physics and 
Astronomy, The University of North Carolina at Chapel Hill, Chapel Hill, NC 
27599 \\
$^{2}$National High Magnetic Field Laboratory, 
Florida State University, Tallahassee, FL 32310 \\ 
}
\date{\today}
\maketitle

\begin{abstract}
We explore the scaling description for a two-dimensional 
metal-insulator transition (MIT) of electrons in silicon.  
Near the MIT, $\beta_{T}/p = (-1/p)d(\ln g)/d(\ln T)$ 
is universal (with $p$, a sample dependent exponent, determined
separately; $g$--conductance, $T$--temperature).  We obtain the 
characteristic temperatures $T_0$ and
$T_1$ demarking respectively the quantum critical region and the
regime of validity of single parameter scaling in the metallic phase,
and show that $T_1$ vanishes as the transition is approached.  For
$T<T_1$, the scaling of the data requires a second parameter.
Moreover, all of the data can be described with two-parameter scaling 
at all densities -- even far from the transition.

\end{abstract}

%
%
%
]

Many recent experiments have provided strong evidence for a two-dimensional
(2D) disordered metallic state in Si metal-oxide-semiconductor field-effect
transistors (MOSFETs) and in other 2D 
systems~\cite{Krav,DP_PRL,pGaAs,pSiGe,Pudalov1,Shayegan,nGaAs},
but the microscopic nature of the metallic phase still awaits a solid 
understanding.  Since there have been predictions for 2D metallic behavior for
quite some time~\cite{Finkelstein}, it should not be too surprising to see a 
metal-insulator transition (MIT), but on the other hand it appears to violate 
rather general and widely accepted assumptions~\cite{gang}. It is apparent 
from the experimental data~\cite{DP_PRL,Pudalov1,Kravbpar,Kawaji,spinflip}
that spin interactions are an important component of the metallic 
behavior~\cite{Finkelstein,newgang,triplet}, and it is suspected that the 
dynamical Coulomb interactions are 
important~\cite{Finkelstein,newgang,triplet} since the non-interacting
(screened) carriers are known to be insulating in 2D~\cite{gang}.

The physics of the MIT can be reduced to certain scaling forms that describe 
the behavior of the conductivity $\sigma$ as a function of temperature $T$, 
sample length $L$, and carrier density $n$.  A simple single-parameter scaling
has proved to be a successful description of the critical behavior of $\sigma$
in most 2D systems.  It fails, however, in the metallic phase at the lowest 
temperatures, where $\sigma (T)$ becomes very weak and, in some systems, it 
even appears to saturate as $T\rightarrow 0$.  From the careful analysis of
the experiments discussed below, we find that two-parameter scaling is more 
appropriate in that regime.  In fact, we show that {\em all of the data} on 
both insulating and metallic sides of the transition can be described with 
two-parameter scaling at all densities -- even far from the transition.  Such 
scaling has been proposed theoretically near the MIT in the presence of 
dangerously irrelevant variables\cite{two-param}, and it also provides an 
excellent description of the MIT in a different type of the 2D 
metal~\cite{Kondo1}.  This form of the two-parameter scaling function is 
consistent with the apparent success of the conventional single-parameter 
scaling in the limited range of (n,T) phase space.

One signature of the 2D MIT is a reversal of the sign of $d\sigma/dT$.  
($\sigma$ is written throughout dimensionlessly -- so all numerical values are
multiplied implicitly by $e^2/h$ to recover SI units).  Measurements of the 
conductance $G=(w/L)\sigma$ ($w$ is device width) as a function of the carrier
density $n$ at several values of $T$ reveal a crossing point at a critical 
density $n_c$.  Near enough to the MIT, a one-parameter scaling scheme holds,
\begin{equation}
\sigma(n,T) \sim f(\dnb) 
= f(\delta_n/ T^{p}) ,
\label{scale1}
\end{equation}
where $\delta_n = (n-n_c)/n_c$, $\dnb =\dn/T^p$,
and $p = 1/z\nu$ is the scaling exponent describing the critical
behavior of correlations scales of temporal and spatial fluctuations.
$f(\dn/T^{p})\sim \exp(\delta_n/ T^{p})$ for $T > T_{0}\sim |\dn|^{1/p}$ 
in the quantum critical region~\cite{newgang}.  Even for $T < T_0$, the data 
still can be scaled according to Eq.~(\ref{scale1}) to form a single curve.  
The single-parameter scaling fails at $T=T_{1} < T_0$, and we show that the 
scale $T_{1}= T_{1}(\dn)$ vanishes as the transition is approached.

We examine the phase diagram of $\sigma$ near the 2D MIT in two samples with 
peak mobility $\mu_{4.2} \simeq 0.5$~m$^2$/V sec at $T=4.2$~K.  To obtain 
$T<4.2$~K, we used a $He^3$-$He^4$ dilution refrigerator with heavily shielded
wiring.  Small AC signals were measured with lock-in amplifiers. The samples 
were generic two-probe Si MOSFETs with a 50~nm thick oxide layer; sample 17
was $L \times w = 5 \times 11.5~\mu$m and sample 27 was $254 \times 254~\mu$m.
Both devices have a density of oxide interface states $N_{ox}$ \lap 
$10^{14}/$m$^2$, which is measured with standard techniques~\cite{AFS}.

\begin{figure}
\epsfxsize=3.2in
\epsffile{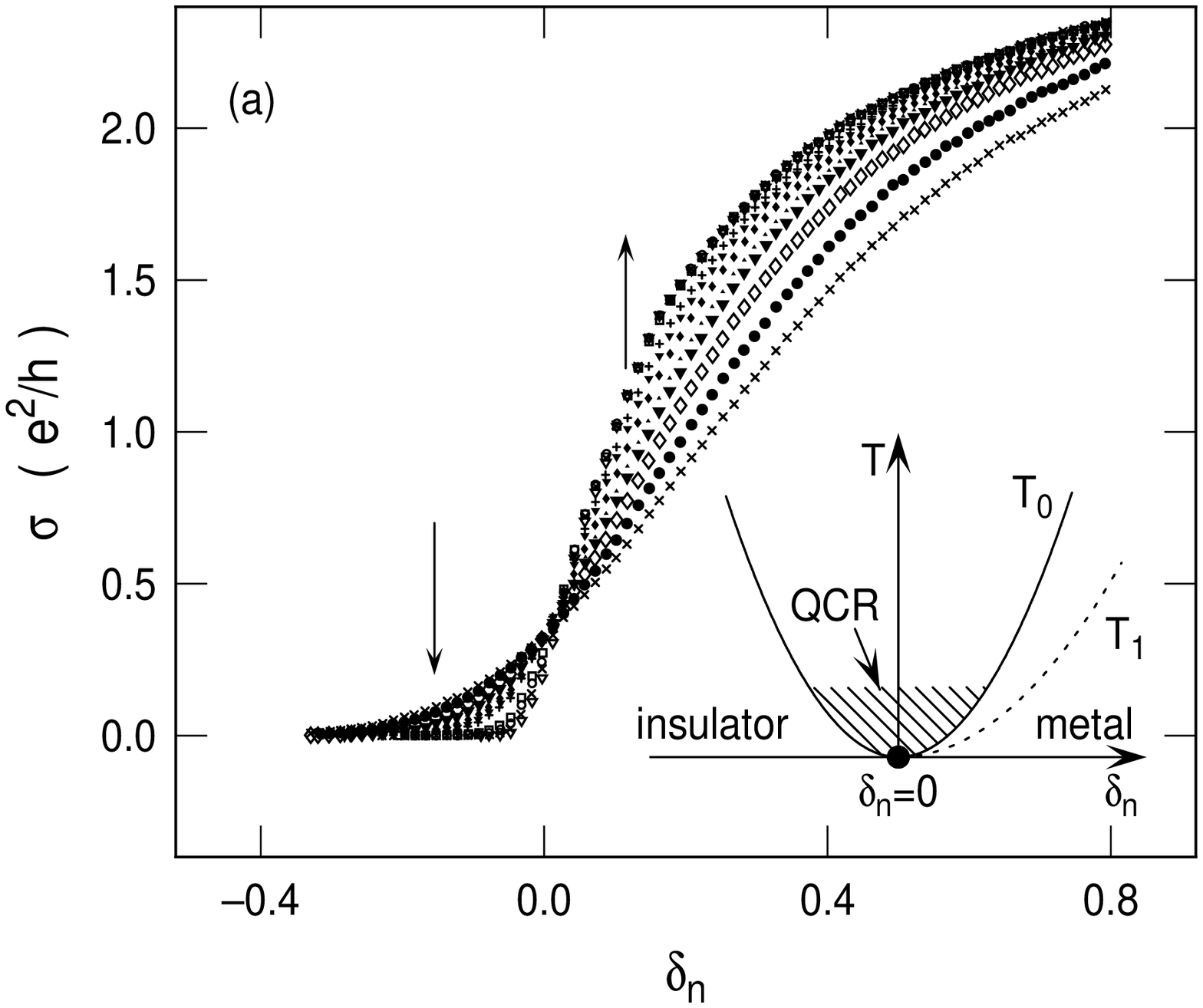}
\vglue 0.1 cm
\epsfxsize=3.2in
\epsffile{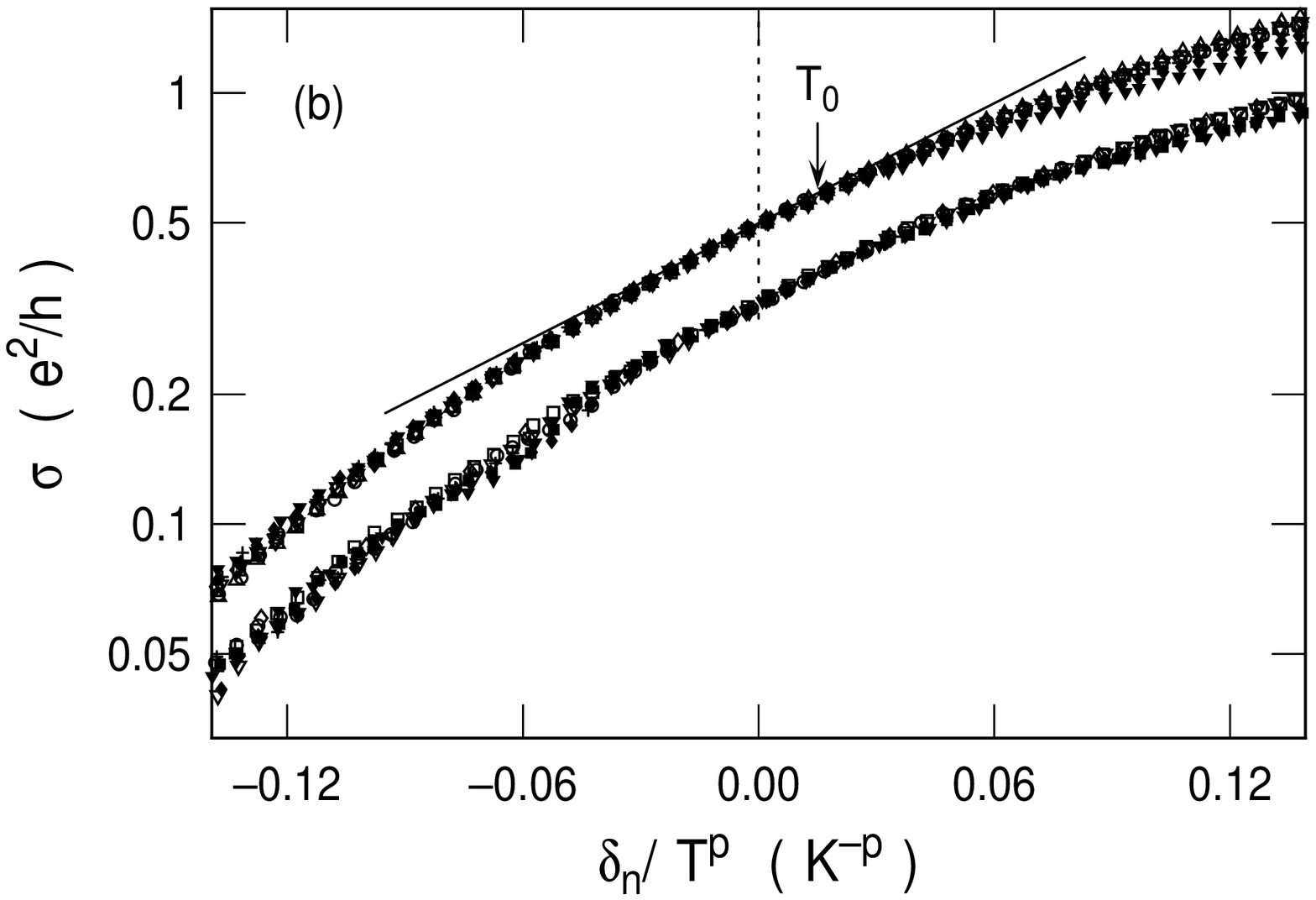}
\vglue 0.1 cm
\epsfxsize=3.2in
\epsffile{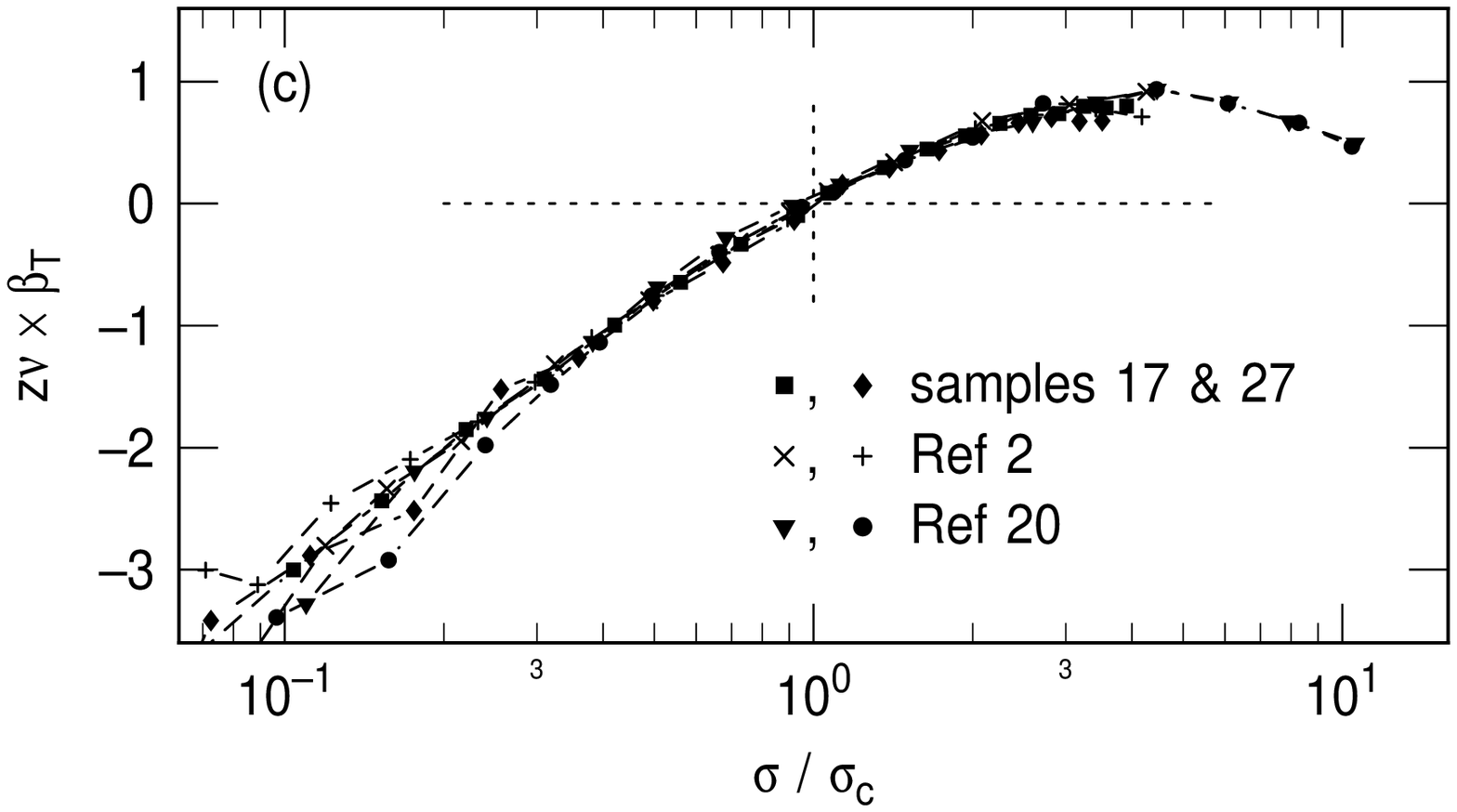}
\vglue 0.1 cm
\caption{
(a) $\sigma(\dn,T)$ for 0.4~K $ < T < $ 4~K from sample 17 with a schematic 
phase diagram inset.  Arrows in the main figure indicate the direction of 
flow for decreasing temperature.  
The phase diagram comprises two curves for $T_0 \propto |\dn|^{1/p}$ 
(solid line) and $T_1 \propto |\dn|^{1/q}$
(dotted line) that separate the $\delta_n - T$ phase space into three 
regions.  The quantum critical regime (QCR) is indicated schematically by 
cross-hatching.  For $T > T_1$ a single parameter can be used to scale 
the conductivity and for $T < T_1$ scaling with one parameter fails.
(b) Scaling of data in $\ln \sigma$ {\it vs} $\delta_n/T^p$ 
for both samples (upper curve is sample 27 and lower curve is sample 17).  
The solid line emphasizes the linearity of $\ln \sigma$ in $\dn$ near the
MIT.  
(c) In the regime of validity of one-parameter scaling ($T>T_1$), the 
universal behavior of $z\nu\beta_T\sim\nu\beta(g)$ is shown for six
Si MOSFETs that span a wide range of parameters.
}
\label {phdiag}
\end{figure}

\begin{figure}
\epsfxsize=3.2in
\epsffile{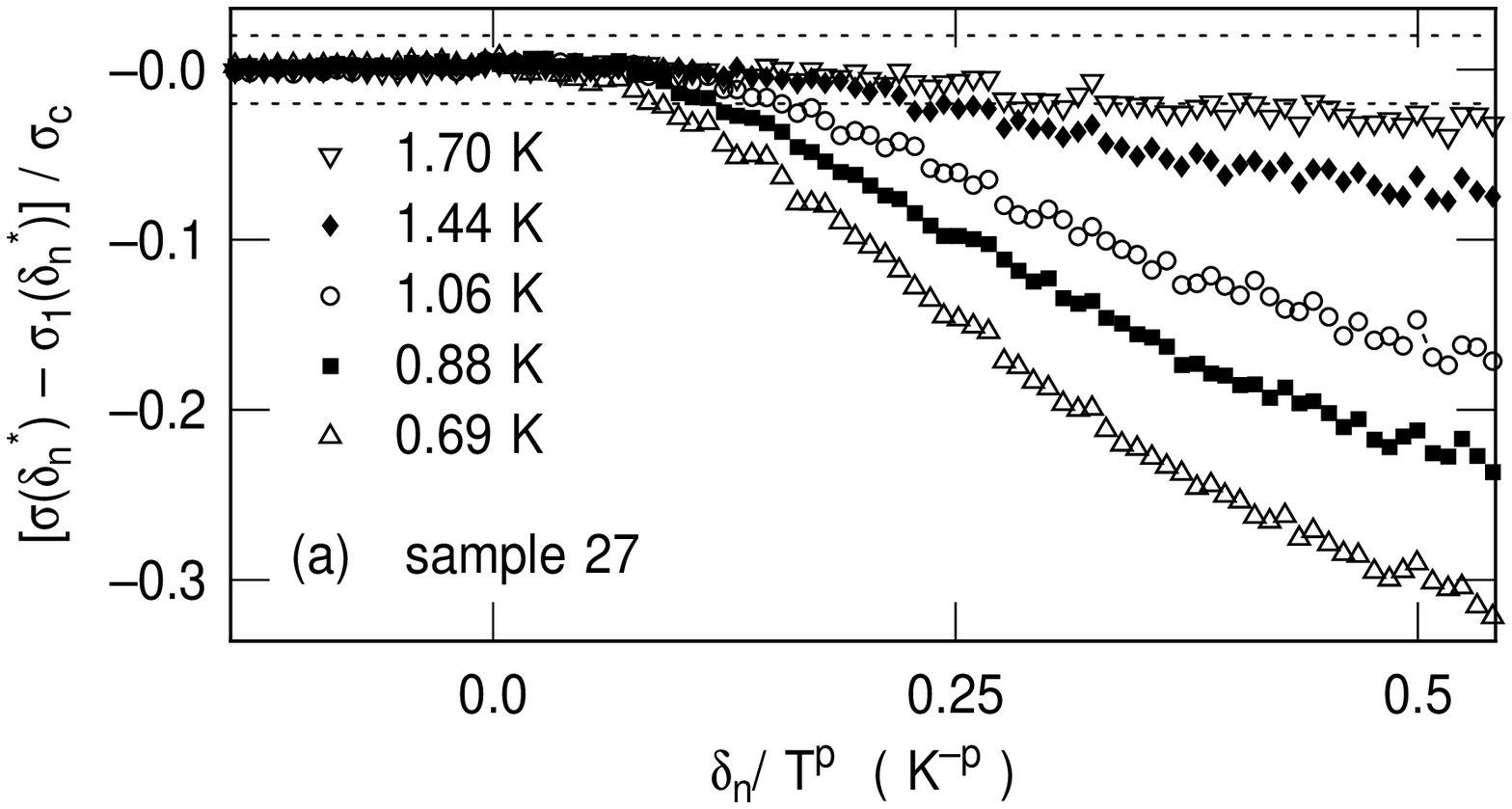}
\vglue 0.1 cm
\epsfxsize=3.2in
\epsffile{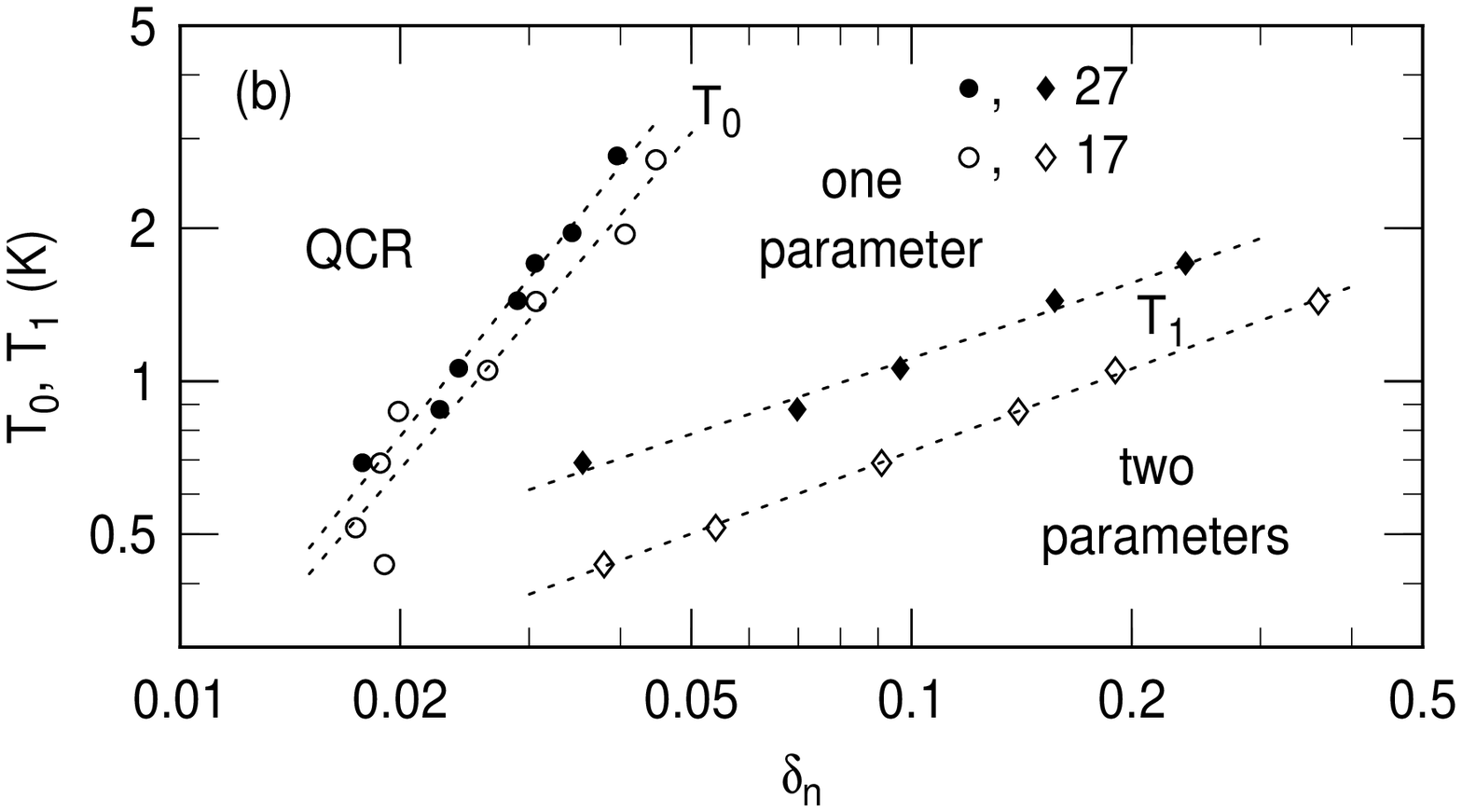}
\vglue 0.1 cm
\caption{
(a) Method of observing crossover from one-parameter to 
two-parameter regimes by measuring deviations from the $\sigma_1$ at each
$T$.  Dotted lines depict (arbitrarily chosen) 2\% 
deviation.  
(b) Experimental phase diagram for both samples.  Open and solid dots 
mark $T_0$, open and solid diamonds mark $T_1$.  For each sample, the 
phase diagram is separated into the quantum critical regime (QCR)
where $\ln \sigma \propto \dnb$, 
$T_{1} < T < T_0$ regime where the T-dependence of $\sigma$ is weaker but
the one-parameter scaling still works, and the 
$T<T_1$ regime where scaling can be achieved only with two parameters.
}
\label {T1}
\end{figure}

Figure~\ref{phdiag}(a) contains $\sigma(\dn,T)$ from a 2D inversion layer
(sample 17) below 4 K, and the conventional one-parameter 
scaling~\cite{newgang} results appear in Fig.~\ref{phdiag}(b).  The individual
curves of $\sigma(\dn)$ at fixed $T$ all cross at a critical density $n_c$ as 
expected.  As shown in Fig.~\ref{phdiag}(b) the raw data for $\sigma(\dnb)$
collapse on to a single curve [``master curve'' 
$\sigma_{1}(\dnb)\sim f(\dnb)$] for a given 2D system if a certain exponent 
$p$ is chosen in $\dnb = \dn/T^p$.  This collapse fits the suggestion of a 
quantum critical point at $T = 0$ and $\dn = 0$.  Critical conductivities in 
other systems fall in the range $0.3 < \sigma_c < 5$, and so do the present 
results.  Sample 27 exhibits the same features as sample 17 with slightly 
different parameters: $\sigma_c$ of the two samples (0.32 for sample 17 and 
0.50 for sample 27) differ from each other by 50\% even though the values of 
$n_c$ ($1.7 \times 10^{15}/$m$^2$ and  $1.9 \times 10^{15}/$m$^2$, 
respectively) and the exponents $p$ (1/2.0 and 1/2.2, respectively) are within
10\%.  As predicted\cite{newgang}, the scaled data $\ln \sigma(\dnb)$ in 
Fig.~\ref{phdiag}(b) are 
approximately 
linear near enough to the transition, for $T>T_0$ 
(emphasized by the solid line).  This is tantamount to the symmetry between
$\sigma(\dn)$ and $1/\sigma(-\dn)$, which has been predicted~\cite{newgang}
and observed~\cite{refl-symm}.  Beyond the linear range (marked in the figure
as $T_0$), the data continue to scale on both the insulating and metallic 
sides.  At some point, however, the scaling fails on the metallic side and the
individual traces of $\sigma(\dnb)$ at fixed $T$ diverge from 
$\sigma_1(\dnb)$.  These points in $\dnb$ can be read as positions in the 
phase diagram for the MIT.  A phase diagram like that proposed for quantum 
phase transitions~\cite{QPT} is inset in Fig.~\ref{phdiag}(a).  The critical
point is at $T=0$ and $\dn =0$ and crossover lines separate the quantum 
critical region from the insulating and metallic regions.  We have added a
second (dotted) curve $T_1 (\dn)$ demarking the region ($T < T_1$) where the 
conventional one-parameter scaling breaks down.  $T_1$ was obtained from 
individual traces of $\sigma(\dnb)$ by locating their departure from 
$\sigma_1(\dnb)$ as shown in Fig.~\ref{T1}(a).  It is obvious that the 
difference between one-parameter scaling and the experiment can reach as 
much as $[\sigma(\dnb) - \sigma_1(\dnb)]/\sigma_c =$ 35\%.
The $T_1$ are plotted as a function of $\delta_n$ in Fig.~\ref{T1}(b),
which shows clearly that the higher temperature data can be scaled over a 
wider range of $\delta_n$.  From Fig.~\ref{T1}(b), the characteristic 
temperature $T_1$ appears to have a power-law dependence on $\delta_n$, as 
might be expected near a critical point.  The power-law dependence of $T_1$ 
obtains in both samples and the exponents are very similar: 0.54 $\pm$ 0.01 
for sample 17 and 0.49 $\pm$ 0.03 for sample 27.

For $T > T_1$, one-parameter scaling holds.  We show that, even though 
different parameters $p$, $n_c$ and $\sigma_c$ are required to obtain the 
master curves, there is a universal feature to the ultimate scaling function 
of the conductance as a function of length.  In particular, we find 
that the function $\beta_{T}/p = (1/p)[- d(\ln g)/d(\ln T)]$ is universal.  
The results of this transformation are shown in Fig.~\ref{phdiag}(c) for the 
two samples studied here and compared with four other samples studied in the
past~\cite{DP_PRL,KPLi}.  Clearly the curve 
$\beta_T/p$ is universal in the vicinity of the MIT.  These experiments span 
two orders of magnitude in sample dimension~\cite{KPLi}, an order of magnitude
in critical exponent $p$~\cite{KPLi}, about an order of magnitude in 
$\mu_{4.2}$, different surface electrostatics~\cite{DP_PRL}, and different
manufacturing procedures.  We expect the relation to hold in other 2D systems,
and in fact a somewhat similar curve was obtained for higher peak-mobility Si
MOSFETs~\cite{Furneaux}.
With the assumption that the length over which quantum coherence obtains is 
growing algebraically as $T \rightarrow 0$ so that 
$\Lphi \propto 1/T^{1/z}$~\cite{thouless}, we can write the scaling 
law~\cite{gang} $\beta(g) = d(\ln g)/d(\ln L) \sim d(\ln g)/d(\ln \Lphi )$ as 
$\beta(g)\sim z\beta_T$.  Fig.~\ref{phdiag}(c) thus shows the universality of
$z\nu\beta_T\sim\nu\beta(g)$.  Recent numerical work has 
suggested~\cite{universal_beta} that $\nu\beta(g)$ is a universal function
close to the MIT.

\begin{figure}
\epsfxsize=3.3in
\epsffile{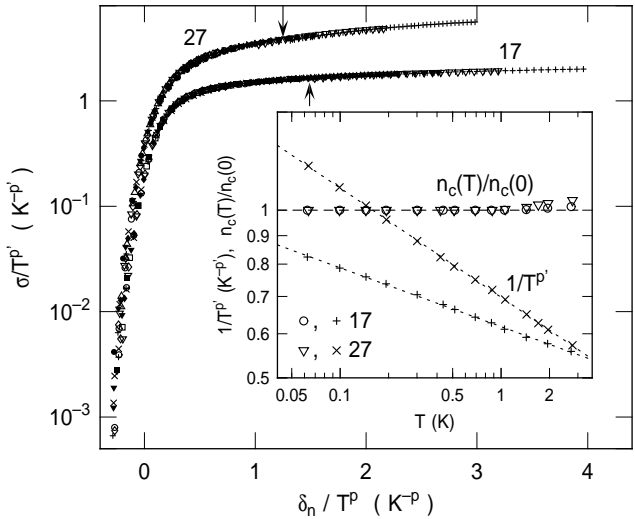}
\vglue 0.1 cm
\caption{
Two-parameter scaling of {\em all of the data} for both samples according
to Eq.~(2).  The parameters from the new scaling law are inset.  The critical 
concentration $n_c$ is compared to its low-temperature value to show the small
change at higher temperature.  For $T < 0.4K$ only limited ranges of $\dn$ can
be scaled (see text).  The arrows mark the end of the scaling curves if all 
data below 0.4~K are removed.  
}
\label {newscal}
\end{figure}

On the metallic side, for $T < T_1$, we find that each $\sigma (\dnb)$ trace 
deviates from the master curve in the same way, regardless of temperature, 
which suggests that it should be possible to scale all of the curves using two
parameters.  This, together with the predictions of two-parameter scaling for 
certain forms of MIT ({\it e.g.} with a strong triplet coupling, {\it i.e.} 
the spin-dependent part of the electron-electron interaction is 
large)~\cite{two-param}, encouraged us to examine a new scaling scheme 
according to
\begin{equation}
\sigma (n,T)\sim T^{p'}f(\delta_{n}/T^{p}).
\label{2scaling}
\end{equation}
This scaling form has been applied to a different form of the 2D metal with 
considerable success~\cite{Kondo1}.  As can be seen in Fig.~\ref{newscal}, 
with this form we can scale {\it all of the data} for $0.4 < T < 1.5$~K from 
the insulating side through to the metallic side using the {\em same} values 
of $p$ that were used to accomplish single-parameter scaling at $T > T_1$.  
For each sample, the data collapse onto a single (sample dependent) curve over
a range of concentrations $\dn$ far larger than possible with the 
one-parameter scheme above.  At $T=0.4$~K, $k_{B}T\approx 0.02E_F$ ($E_F$ -- 
Fermi energy) at the MIT, {\it i.~e.} lower than in many other MIT experiments
on non-silicon systems.  In addition, Fig.~\ref{newscal} shows that the data 
obtained at even lower $T$ ($0.06 < T < 0.4$~K) follow the same scaling law 
(Eq.~\ref{2scaling}).  For $\dn < 0.2$, there is a maximum in $\sigma (T)$ at
$T=T_m$ and the scaling fails.  This maximum has been 
attributed~\cite{spinflip} to Kondo coupling to disorder-induced local 
magnetic moments.  The local moments did not affect the transport at 
$T_m < T$, consistent with our finding (Fig.~\ref{newscal}) that all of the 
data for $T_m < T$ follow the {\em same} two-parameter scaling form 
(Eq.~\ref{2scaling}).  Furthermore, this same two-parameter scaling holds even
at $T> 1.5$~K if we allow a small variation ($<5 \%$) of $n_c$ with 
temperature (see Fig.~\ref{newscal} inset).  This small change in $n_c$ might 
result from rescaled screening or other finite temperature effects~\cite{ncT}.

The power-law dependence of the prefactor for $\sigma$ on $T$ is shown
in the inset of Fig.~\ref{newscal}.  The new exponent $p'$ appears
with values $0.1$ for sample 17 and $0.2$ for sample 27.  Clearly,
such a weak temperature dependence of the prefactor can become
observable only when the scaling function $f(\dnb)$ itself becomes a
very weak function of $T$, which happens at some $T\approx T_{1} <
T_0$ in the metallic regime.  Indeed, Fig.~\ref{phdiag}(c)
($z\nu\beta_T$) shows that, for $\sigma/\sigma_c$ \gap 4 in the
metallic phase, $f(\dnb)$ becomes a power-law function of $T$ with an
exponent comparable to the values of $p'$ found from the two-parameter
scaling.  This explains naturally the apparent success of the
single-parameter scaling in the description of the 2D MIT, and the
existence of the energy scale $T_1 (\dn)$ where it fails.  The two
samples, which differ in length by one and a half orders of magnitude,
follow the scaling form (\ref{2scaling}) albeit with somewhat
different exponents $p'$.  It has been suggested~\cite{KPLi} that
finite size effects may play a role in the failure of one-parameter
scaling.  On the other hand, a larger value of $p'$, {\it i.e.} a
larger deviation from one-parameter scaling, is found in a larger
sample, but it would be interesting to study if there is any
systematic dependence of all the critical exponents on the sample
length.

The data (Fig.~\ref{newscal}) also show that $\sigma_c$ depends very weakly on
temperature: $\sigma_{c}=\sigma(\dn=0,T)\propto T^{p'}$, resulting in 
$\sigma_{c}=0$ at $T=0$.  It is important to note, however, that the direct 
observation of a significant reduction of $\sigma_c$ as $T \rightarrow 0$ can 
not be achieved at experimentally accessible temperatures because of the very 
small exponent $p'$.  For this reason, our result is not inconsistent with 
the constant values of $\sigma_c$ observed in other experiments.  Of course, 
we also can not distinguish this temperature dependence from a logarithm, 
which is another form proposed for corrections to scaling for certain quantum 
critical points~\cite{logscaling}.  On the other hand, our result is in
agreement with the recent work~\cite{Kondo1} on the transition between a 
different type of the 2D metal and an insulator, but the values of $p'$ are 
considerably different, presumably reflecting the different universality 
classes of the two situations.

In the metallic phase, as $T \rightarrow 0$, $\sigma (T)$ becomes a very weak 
function after an initial rapid increase.  As a result of this 
``saturation''~\cite{saturation} of $\sigma (T)$, single-parameter scaling 
fails at the lowest temperatures (see also Ref.~\cite{Pudalov2}) and the scale 
$T_1$ must exist.  We have identified explicitly $T_{1}=T_{1}(\dn)$ and shown 
that it vanishes as the transition is approached.  Furthermore, we have shown 
that {\em all} of the conductivity data in both metallic and insulating 
regimes can be scaled with two parameters according to Eq.~(\ref{2scaling}).  
That scaling form is consistent with the observation that $\sigma (T)$ 
becomes very weak (``saturates'') in the metallic phase as $T\rightarrow 0$.  
We reiterate that the scaling form (\ref{2scaling}) has been predicted for 
certain models \cite{two-param}.  The predictions do not, however, decribe our
data in detail ({\it e.g.} the sign or magnitude of $p'$).  Scaling 
form~(\ref{2scaling}) has been also used successfully to describe a different
type of a 2D MIT~\cite{Kondo1}.

The authors are grateful to K.P.\ Li for technical assistance, V.\ 
Dobrosavljevi\'{c} for useful discussions, and to the Aspen Center for Physics
for hospitality.  This work was supported by NSF Grant No. DMR-9796339.


\begin{references}
\bibitem{ncT} $T$-dependent critical density was also observed in
another experiment~\cite{Hanein-pcT}, where the crossing point
was well-defined (independent of $T$) only below $T\approx 0.150$~K.  This
saturation $T$ is about ten times lower than in our experiment
($\approx 1.5$~K) consistent with the Fermi energy at the MIT being about
ten times smaller than in our samples.
\bibitem{saturation} Our results and data from higher peak-mobility Si 
MOSFETs~\cite{Pudalov2} show that, in the metallic phase, $\sigma$ continues 
to increase as $T\rightarrow 0$.
\end{references}
\end{document}